\documentclass[12pt]{article}
\usepackage[a4paper]{geometry}
\usepackage[english]{babel}
\usepackage{tikz}
\usepackage{tikz-cd}
\usepackage{mathrsfs}
\usepackage{amsmath}
\usepackage{amssymb}

\newcommand{\1}{\mathbf{1}}

\newcommand{\G}{\mathcal{G}}

\usepackage[pagebackref,draft=false]{hyperref}
\hypersetup{colorlinks,
linkcolor=myrefcolor,
citecolor=mycitecolor,
urlcolor=myurlcolor}

\usepackage[capitalize]{cleveref}
\usepackage{caption}
\usepackage{etaremune}

\usepackage{xcolor}
\definecolor{myurlcolor}{rgb}{0,0,0.4}
\definecolor{mycitecolor}{rgb}{0,0.5,0}
\definecolor{myrefcolor}{rgb}{0.5,0,0}
\usepackage{graphicx}
\usepackage{tikz}
\usepackage{tikz-cd}
\usepackage{mathrsfs}

\usepackage{etoolbox}
\usepackage{makeidx}
\usepackage{sectsty}
\usepackage{enumitem} 
\usepackage{latexsym}
\usepackage{braket}
\usepackage{caption}
\usepackage[utf8]{inputenx}
\usepackage[T1]{fontenc}
\usepackage{lmodern}
\usepackage{textcomp}
\usepackage{microtype}
\usepackage{totcount}
\usepackage{blindtext}

\newcommand{\be}{\begin{equation}}
\newcommand{\ee}{\end{equation}}

\author{F. M. Ciaglia$^{1,6}$  \href{https://orcid.org/0000-0002-8987-1181}{\includegraphics[scale=0.7]{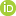}}, F. Di Cosmo$^{2,3,7}$\href{https://orcid.org/0000-0003-0256-5913}{\includegraphics[scale=0.7]{ORCID.png}}, A. Ibort$^{2,3,8}$\href{https://orcid.org/0000-0002-0580-5858}{\includegraphics[scale=0.7]{ORCID.png}}, G. Marmo$^{4,5,9}$\href{https://orcid.org/0000-0003-2662-2193}{\includegraphics[scale=0.7]{ORCID.png}} \\
\footnotesize{$^{1}$\textit{ Max Planck Institute for Mathematics in the Sciences, Leipzig, Germany}} \\
\footnotesize{$^{2}$\textit{ ICMAT, Instituto de Ciencias Matem\'{a}ticas (CSIC-UAM-UC3M-UCM)}}  \\
\footnotesize{$^{3}$\textit{ Depto. de Matem\'aticas, Univ. Carlos III de Madrid, Legan\'es, Madrid, Spain}}  \\
\footnotesize{$^{4}$\textit{ Dipartimento di Fisica ``E. Pancini'', Universit\`a di Napoli Federico II, Napoli, Italy}} \\
\footnotesize{$^{5}$\textit{ INFN-Sezione di Napoli, Napoli, Italy.}} \\
\footnotesize{$^{6}$\textit{ e-mail: \texttt{florio.m.ciaglia[at]gmail.com}}, $^{7}$\textit{ e-mail: \texttt{fcosmo[at]math.uc3m.es}}, } \\
\footnotesize{$^{8}$\textit{ e-mail: \texttt{albertoi[at]math.uc3m.es}}, $^{9}$\textit{ e-mail: \texttt{marmo[at]na.infn.it}}}
}

\date{}

\begin{document}

%

\title{Schwinger's picture of Quantum Mechanics}

\maketitle

\begin{abstract}
In this paper we will present tha main features of what can be called Schwinger's foundational approach to Quantum Mechanics. The basic ingredients of this formulation are the \textit{selective measurements}, whose algebraic composition rules define a mathematical structure called groupoid, which is associated with any physical system. After the introduction of the basic axioms of a groupoid, the concepts of observables and states, statistical interpretation and evolution are derived. An example is finally introduced to support the theoretical description of this approach.  
\end{abstract}



\section{Introduction}

At the dawn of modern Quantum Mechanics, the founding fathers faced two pressing and contradicting evidences. On one hand the atomistic/discrete structure of matter, which was evident in the structure of spectroscopic lines with its peculiar combination relations given by the Rydberg-Ritz composition rule and, on the other hand, with the statistical/probabilitic behaviour emerging in scattering problems, complicated even more with interference phenomena (previous to the two-slit experiment were Davisson-Germer experiment on the diffraction of electrons scattered by a surface of a crystal of nickel metal \cite{Da28,Ma04}). 

The spectroscopic phenomenology gave rise to matrix mechanics (mostly elaborated by Heisenberg, together with Jordan and Born), while the interpretation of matter waves gave rise to wave mechanics (which may be ascribed to de Broglie and Schr\"{o}dinger). An additional nuisance came with the doubling of some spectral lines and the Stern-Gerlach experiment. They led to the introduction of a specific ``quantum'' entity, the spin. The appearance of phenomenology due to the presence of spinning particles gave rise first to Pauli's equation and, eventually, to the triumph of Dirac's equation, which replaced the non-relativistic Schr\"{o}dinger's equation. 

This dualistic aspect of particles and waves was synthetized by the complementary principle of Bohr and was represented by the dialectic aspect of ``being and becoming'' (the indetermination relation between position and momentum). Even though, as it was shown by Schr\"{o}dinger and von Neumann, the two pictures were equivalent; Schr\"{o}dinger's wave mechanics and Heisenberg's matrix mechanics rested on alternative formulations. The atomistic/discrete aspect was prevalent in the second one while the statistical-probabilistic aspect was prevalent in the first one, the matter-wave picture. 

Almost twenty-five years after this golden-age period, J.Schwinger took up the foundational aspects of Quantum Mechanics with his proposal of ``Symbolism of atomic measurements''. Schwinger's positivistic attitude is well described using his own words: ``\textit{... we have to recognize the fundamental philosophical conception that Physics is an experimental science; it is concerned only with those statements which in some sense can be verified by an experiment. The purpose of the theory is to provide a unification, a codification, or however you want to say it, of those results which can be tested by means of some experiment. Therefore, what is fundamental to any theory of a specific department of nature is the theory of a measurement within that domain}'' (see \cite[p.10]{Sc91}  Quantum Mechanics. Symbolism of Atomic Measurements).

In modern mathematical terms we could say that the description of a physical system in the Schr\"{o}dinger picture starts by associating with any system a complex separable Hilbert space, typically the space of square integrable functions on some domain whose vectors are interpreted as the ``waves'' (wave functions) of the theory, while in the Heisenberg picture we start by associating with every system an algebra (a $C^*$-algebra in more technical words) of the observables of the system. In this short paper we would like to argue that in the Schwinger picture we associate with any physical system a groupoid that describes what Schwinger called \textit{selective measurements} and whose determinating relations (composition rules) provide the symbolic algebra of quantities adequate to describe atomic phenomena: ``\textit{The laws of atomic physics, must be expressed, therefore, in a non-classical mathematical language that constitutes a symbolic expression of the properties of microscopic measurements}'' (see \cite{Sc91} Quantum Kinematics and Dynamics). 

Thus, on elaborating on the algebra of measurements as introduced by Schwinger, we will show that similarly to the Hilbert space or the $C^*$-algebra of Schr\"{o}dinger and Heisenberg pictures, respectively, the Schwinger picture starts with a groupoid. Out of it, we should exhibit: observables, states, probability/statistical interpretation, evolution and a rule of composition of systems. Of course, we will not be able to deal with all these aspects in any detail and we have to refer to our previously published papers (see \cite{Ci18,Ib18,Ib18b,Ib18c}). However, we shall outline how the groupoid picture is able to provide an alternative picture to those of Heisenberg and Schr\"{o}dinger, which may be called a Schwinger picture of Quantum Mechanics. 

\section{The algebra of selective measurements and groupoids}

In setting up a theory, a number of primitive notions have to be chosen. ``Primitive'' here does not imply that they cannot be scrutinized, just the contrary, a formal justification must be provided for them, and its consistency and capability have to be assessed. For instance, in the Schr\"{o}dinger picture of Quantum Mechanics the mathematical structures used in the theory are ``primitive'' in this sense. The notion of Hilbert spaces, vectors and operators are introduced and freely used in the development of the theory. Even more important, the epistemological content of the theory is determined in this way. The primitive notions of ``observables'' and ``states'' are then established. The former as those quantities characteristic of the system that can be ``observed'' as well as the axiom identifying them with self-adjoint operators in a given Hilbert space, and the latter as the given system once a maximal family of compatible measurements have been performed on it, i.e., once that a maximal amount of information has been obtained from it. An additional axiom is introduced in this picture identifying such pure states with rays in a given Hilbert space. 

Schwinger's approach is widely divergent with those primitive assumptions and it focuses on the algebraic properties of certain elementary measurement processes, called ``\textit{selective measurements}''. This approach is reminiscent of the later developed ``propositional calculus'' approach to Quantum Mechanics (\cite{BvN36, Ja68}). We do not, however, proceed in this direction and focus solely on Schwinger's algebra of selective measurements. In Schwinger's own words a (generalized) selective measurement is ``\textit{a device that selects the system (in a given ensemble) with well defined values $a$ for the observable $A$, and returns it with well defined values $b$ for the observable $B$}''. We will denote such generalized measurement $M(b,a)$ and it describes a process by which the specific system under study suffers a change in its properties as indicated above. Such change will be caused either by the interaction with the given external device or because of its own dynamical behaviour. We can consider for instance an excited state of an atom. Such system could suffer a transition to a lower energy level emitting a photon, or it may be excited to a higher energy level by illuminating it with a laser beam of the appropriate frequency.   

We have chosen to consider these changes in the physical properties of a given system as primitive concepts. They will be called in what follows ``\textit{transitions}'' (elementary transitions if they cannot be decomposed further into other transitions caused by similar devices). Thus transitions will constitute an abstraction of Schwinger's \textit{selective measurements} and they can be thought as the transitions amplitudes among energy levels or the transformations suffered by the spin of an electron when it passes through a Stern-Gerlach device. 

Restricting our attention to selective measurements $M(a',a)$ where $a'$, $a$ are outcomes of the same family of compatible observables $\mathbb{A}$, then it is clear that there is a natural composition law $M(a''',a'')\circ M(a',a)$ that consists in ``selecting'' the system first by means of $M(a',a)$ and immediately afterwards with $M(a''',a'')$. But then the composition $M(a''',a'')\circ M(a',a)= M(a''',a)$ will be defined only if $a''=a'$, because otherwise no system will be selected, and the result will be $M(a''',a)$. Then if we call $a$ the ``source'' of $M(a',a)$ and $a'$ the ``target'' of $M(a',a)$ we will say that two selective measurements $M(a''',a''), \,M(a',a)$ can be composed only if the source of $M(a''',a'')$ is the target of $M(a',a)$. Such composition rule is clearly associative:
\begin{equation}    
M(a''',a'')\circ\left( M(a'',a') \circ M(a',a) \right) = \left( M(a''',a'') \circ M(a'',a') \right) \circ M(a',a)
\label{eq:composition} 
\end{equation}
and it possesses units, i.e. selective measurements $M(a,a)$ such that:
\begin{equation}
M(a',a')\circ M(a',a) = M(a',a)\,,\quad M(a',a)\circ M(a,a) = M(a',a)\,.
\label{eq:units}
\end{equation}
Moreover for any selective measurement $M(a',a)$ we may imagine another device such that it determines a selective measurement $M(a,a')$ that ``undoes'' $M(a',a)$, i.e.:
\begin{equation}
M(a',a)\circ M(a,a')= M(a',a')\,; \quad M(a,a')\circ M(a',a) = M(a,a)\,.
\label{eq:inverses}
\end{equation}
The previous list of axioms, Eq.\eqref{eq:composition}-\eqref{eq:inverses}, defines what is called a groupoid structure on the collection of selective measurements associated with the family of compatible observables $\mathbb{A}$. We will use these ideas as a departing point for our conceptualization of the description of physical systems by abstracting them and stating:

A physical system is described by a groupoid $\G \rightrightarrows \Omega$ where its morphisms \hbox{$\alpha\,:\,a\, \rightarrow \,a'$, $a,\,a' \in \Omega$} will be called \textit{transitions} of the system. The objects of the groupoid, i.e. the elements $a \in \Omega$, will be called the \textit{outcomes} of the system. Finally the units corresponding to the outcome $a$ will be denoted by $\1_a$, $a\in \Omega$. The source and the target maps will be defined, respectively, as $s(\alpha)=a $  and $t(\alpha)=a'$, where $\alpha:\,a\,\rightarrow\,a'$ is the transition that changes the system from returning the outcome $a$ to $a'$ when $\mathbb{A}$ is measured. If the transitions $\alpha$, $\beta$, are such that $s(\alpha) = t(\beta)$ they will be said to be composable and their composition will be denoted by $\alpha \circ \beta$.

Some comments on the physical interpretation of the groupoid axioms are in order now. The restriction on the composition of transitions and associativity of the composition, $\alpha \circ (\beta \circ \gamma) = \alpha \circ (\beta \circ \gamma)$, $\forall \alpha, \beta, \gamma$ composable, reflects the causal structure of the experimental setting. Note in this sense that the right to left composition convention relates to the definition of the future in the laboratory where experiments are performed (while reading the composition from left to right will correspond to ``time'' flowing backwards in the given experimental setting).

The invertibility axiom, $\forall \alpha, \exists \alpha^{-1}$ such that $\alpha \circ \alpha^{-1} = \1_{t(\alpha)}$ and $\alpha^{-1}\circ \alpha = \1_{s(\alpha)}$ reflects Feynman's principle of \textit{microscopic reversibility} (see \cite[p. 3]{Fe05}: ``\textit{The fundamental (microscopic) phenomena in nature are symmetrical with respect to the interchange of past and future}''). 

The outcomes $a\in \Omega$ have the meaning of physical occurrences registered by the physical device used in studying the transitions of the system and they determine in a natural way the units of the groupoid, i.e., $\forall a \in \Omega$ there is a transition $\1_a$ that does not affect the system whenever it returns $a$ when $\mathbb{A}$ is measured, that is, such that $\alpha\circ \1_a = \alpha$; $\1_a\circ \alpha = \alpha$ $\forall \alpha :\, a\, \rightarrow \, a'$.

Given a groupoid $\G \rightrightarrows \Omega$ of transitions describing a physical system, a subset $\mathcal{Q}$ of them will be said to generate it if any other transition can be obtained by composing a finite number of transitions in $\mathcal{Q}$. Moreover the structure $\mathcal{Q}\rightrightarrows \Omega$, also called a quiver, will be said to be irreducible if no transition $\alpha \in \mathcal{Q}$ can be written as $\alpha = \beta \circ \gamma$ with $\beta, \gamma \in \mathcal{Q}$. Such transitions $\alpha \in \mathcal{Q}$ will be said to be elementary or indecomposable (see \cite{Sc51,Sc91}:``\textit{...the uncontrollable disturbance attendant upon a measurement implies that the act of measurement is indivisible, that is to say, any attempt to trace the history of a system during a measurement process usually changes the nature of the measurement that is being performed. Hence to conceive of a given selective measurement $M(a',b')$ as a composed measurement is without physical meaning}'').

Thus, in describing a physical system, we can start by obtaining a quiver $\mathcal{Q} \rightrightarrows \Omega$ of elementary transitions and build the groupoid generated by it. We will see later, in a subsequent section, a fair example of this methodology.

In both standard pictures, Heisenberg and Schr\"odinger, determining a state $\rho$ of a quantum system allows us to provide a statistical interpretation for the system.  We may consider, for instance, two pure states $\rho$, $\rho'$,  which in the Schr\"odinger's picture are defined as rays of a Hilbert space $\mathcal{H}$, and will be represented as the projectors $\rho = \rho_{\psi} = \frac{|\psi \rangle \langle \psi |}{\langle \psi | \psi \rangle}$ and $\rho' = \rho_{\phi} = \frac{|\phi \rangle \langle \phi |}{\langle \phi | \phi \rangle}$, with $|\psi \rangle$, $|\phi \rangle$ vectors in $\mathcal{H}$ respectively.   Such projectors can also be read as a ``coordinate-free'' versions of probability densities.

Consider, in addition, the rank one-operator $\rho_{\psi \phi} = \frac{|\psi \rangle \langle \phi |}{\sqrt{\langle \psi | \psi \rangle \langle \phi | \phi \rangle }}$. It is clear that it satisfies:
$$
\rho_{\psi \phi}^{\dagger} \rho_{\psi \phi} = \rho_{\phi}\, , \qquad \rho_{\psi \phi}\rho_{\psi \phi}^{\dagger}  = \rho_{\psi} \, .
$$ 
Therefore, $\rho_{\psi \phi}$ can be interpreted as a operator-valued square-root of the pure states $\rho_{\psi}$ and $\rho_{\phi}$, thus defining a ``coordinate-free'' version of a transition amplitude between events. The same analogy can be extended also to the mixed states case (see Sec.2.3 in \cite{A-F-M-2019}). In the rest of this short paper, then, we will argue that the abstract counterpart of the coordinate-free transition amplitudes are the morphisms of the groupoid $\G$ which is associated with a given physical system, the rank-one operators being specific realizations of these abstract transitions.

\section{Observables and States}\label{sec: observable and states}

Given a groupoid $\G \rightrightarrows \Omega$ describing (so far just the kinematical structure) a physical system, functions defined on it will be good candidates for observables of the system. Actually, we may construct in several ways a $C^*$-algebra of observables $C^*(\G)$ associated with the groupoid. For instance, we may consider $C_c(\G)$, the algebra of compactly supported functions in $\G$ equipped with the convolution product and $C^*(\G) = \overline{C_c(\G)}$, the closure determined by some natural norm. For instance if $\G$ is discrete countable, $C_c(\G)$ will be the set of functions $f:\,\G\, \rightarrow \, \mathbb{C}$ which are different from zero on a finite number of elements of $\G$, i.e., $f= \sum_{\alpha \in \G} f(\alpha)\delta_{\alpha}$, $f(\alpha)=0$ except for a finite number of $\alpha$'s, and the associative product is defined as
\begin{equation}
f \star g = \left( \sum_{\alpha \in \G} f(\alpha)\delta_{\alpha} \right) \star \left( \sum_{\beta \in \G} g(\beta)\delta_{\beta} \right) = \sum_{s(\alpha)=t(\beta)}f(\alpha)g(\beta)\delta_{\alpha\circ \beta}\,.
\end{equation}
Moreover, the adjoint $f^*$ of the function $f$, is defined as $f^* = \overline{f(\alpha)}\delta_{\alpha^{-1}}$ and the closure can be taken with respect to the weak topology on the space of bounded operator on the Hilbert space $\mathcal{L}^2(\G)$ of square integrable functions on $\G$ where the functions $f$ are represented by means of the left-regular representation $(\lambda(f)\psi)(\beta)= \sum_{t(\alpha)=t(\beta)}f(\alpha)\psi(\alpha^{-1}\circ \beta)$.    

In any case the algebra $C^*(\G)$ carries a $C^*$-algebra structure that distinguishes its real (or self-adjoint) elements, i.e. those ones such that $f^* = f$, and that can be identified with physical observables. Note that in such a case $f(\1_x)=f(x)$ can be interpreted as the actual value of the observable $f$ when the system is ``selected'' at $x\in \Omega$ and $f(\alpha)$ will be associated with the transition amplitude of the observable $f$ between the events $x$ and $y$ connected by the morphism $\alpha:\,x\, \rightarrow \, y$. We will consider our $C^*$-algebra to be unital, i.e. $\1 \in C^*(\G)$. 

States of the system, on the other hand, will be determined as positive normalized linear functionals $\rho:\,C^*(\G)\rightarrow\, \mathbb{C}$, i.e. $\rho(f^* f) \geq 0$ and $\rho(\1) = 1$. This entails the natural statistical interpretation of states as providing the expectation values of observables. Indeed, the number $\rho(f)$ will be interpreted as the expectation value of the observable $f$ (note that if $f$ is real, then $\rho(f)$ is a real number, too), more commonly:   
\begin{equation}
\left\langle f \right\rangle_{\rho} = \rho(f) = \sum_{\alpha \in \G } f(\alpha) \rho(\delta_{\alpha})\,.
\label{eq:expectation value}
\end{equation}
Denoting by $\phi_{\rho}: \, \mathcal{G}\, \rightarrow \, \mathbb{C}$ the function defined by the state $\rho$ on $\G$ (again, assuming that the groupoid $\G$ is discrete) as
\be
\phi_{\rho}(\alpha):=\rho(\delta_{\alpha}),
\ee
then $\left\langle f \right\rangle_{\rho} = \sum_{\alpha \in \G} f(\alpha)\phi_{\rho}(\alpha)$ and the expected value of the observable $f$ can be understood as the mean value of the amplitude $f(\alpha)$ over the groupoid $\G$ with respect to the ``distribution'' $\phi_{\rho}$. The function $\phi_{\rho}$ associated to the state $\rho$ is positive definite, that is:
\begin{equation}
\phi_{\rho}\left(\left(\textstyle{\sum_{i=1}^{N}\zeta_{i}\alpha_{i}}\right)^{*}\,\left(\textstyle{\sum_{j=1}^{N}\zeta_{j}\alpha_{j}}\right)\right)=\sum_{i,j=1}^N \overline{\zeta}_i \zeta_j \phi_{\rho}(\alpha_i^{-1}\circ \alpha_j) \geq 0\, , 
\end{equation} 
for all $N\in \mathbb{N}$, $\zeta_i \in \mathbb{C}$, $\alpha_i \in \G$, when the sum is taken over pairs, $\alpha_i, \alpha_j$ with $\alpha_i^{-1}, \alpha_j$ composable. Notice that, if in addition $f$ is a real observable, Eq.\eqref{eq:expectation value} implies $\rho(f)= \overline{\rho(f)}$, then $\phi_{\rho}(\alpha) = \overline{\phi_{\rho}(\alpha^{-1})}$ and $\phi_{\rho}$ itself is an observable. If in addition $\phi_{\rho}(\alpha^{-1})=\phi_{\rho}(\alpha)^{-1}$ we will call this property of the state $\rho$ ``unitarity''. 

The function $\phi_{\rho}$ characterizes the state $\rho$ and allows us to interprete of the state as providing the probability amplitudes of transitions as it will be discussed in the section to follow. 

\section{The statistical interpretation}

 In order to provide a statistical interpretation of Schwinger's foundational approach to Quantum Mechanics that can be compared with those provided by Heisenberg and Schr\"{o}dinger pictures, we need to provide explicit representations of the groupoid describing the system. Fortunately, given a state $\rho$ of the system, this is just natural, as any state $\rho$ provides a canonical representation $\pi_{\rho}:\, C^*(\G) \, \rightarrow \, \mathcal{B}(\mathcal{H}_{\rho})$, called the GNS representation, of the algebra of observables as bounded operators on a Hilbert space $\mathcal{H}_{\rho}$. The GNS Hilbert space $\mathcal{H}_{\rho}$ is constructed as $\mathcal{H}_{\rho} = \overline{C^*(\G)/I_{\rho}}$, where $I_{\rho}=\left\lbrace f\in C^*(\G) \mid \rho(f^*f)\geq 0\right\rbrace$ is called Gel'fand ideal and the GNS representation is explicitly defined by: 
\begin{equation}
\pi_{\rho}(f)(g+I_{\rho}) = f\star g + I_{\rho}.
\end{equation} 
We will denote $\Psi_f = f + I_{\rho}$ the vector in $\mathcal{H}_{\rho}$ associated to the function $f$ defined on the groupoid $\G$. The inner product $\left\langle  \cdot, \cdot \right\rangle$ on $\mathcal{H}_{\rho}$ is defined as:
\begin{equation}
\left\langle \Psi_f, \Psi_g \right\rangle = \rho(a^*\star b)\,.
\end{equation}

For reason that will be clear in a moment, we will focus on states $\rho$ associated to functions $\phi_{\rho}$ possessing the remarkable property that they factorize with respect to the groupoid composition law:
\begin{equation}
\phi_{\rho}(\alpha\circ \beta) = \phi_{\rho}(\alpha)\phi_{\rho}(\beta)
\label{eq:factorizability}
\end{equation} 
for every pair of composable transitions $\alpha$, $\beta$. Note that $\phi_{\rho}$ do not define a representation (neither a character) of the groupoid. It can be shown (see \cite{Ib18c} for details) that functions $\phi:\, \G \, \rightarrow \, \mathbb{C}$ satisfying Eq.$\eqref{eq:factorizability}$ together with unitarity, do actually define states on $C^*(\G)$. 

The GNS representation associated to such states turns out to be deceptively simple. First we compute the Gel'fand ideal $I_{\rho}$ for such states finding out that 

\begin{eqnarray}
\rho(f^{*}\star f)&=\sum_{x\in\Omega}\,\sum_{\gamma,\beta\in G_{+}(x)}\,\overline{f(\beta)}\,f(\gamma)\,\phi_{\rho}(\beta^{-1}\circ\gamma) = \\ 
& = \sum_{x\in\Omega}\,\overline{\sum_{\beta\in G_{+}(x)} f(\beta)\,\phi_{\rho}(\beta)}\,\sum_{\gamma\in G_{+}(x)}f(\gamma)\phi_{\rho}(\gamma). 
\end{eqnarray}
Denoting by $\Psi_{f}$ the vector associated with $f\in C^{*}(\G)$ as before:
\be
\Psi_f = \sum_{x\in \Omega}\,\Psi_{f}(x)|x\rangle=\sum_{x\in\Omega}\left(\sum_{\alpha\in G_{+}(x)}f(\alpha)\,\phi_{\rho}(\alpha)\right)|x\rangle,
\ee
we get $\rho(f^{*}\,\star f)=||\Psi_{f}||^{2}$, then, $I_{\rho}=\{f\in C^{*}(\G)|\Psi_{f}=0\}$ and $C^{*}(\G)/I_{\rho}$ can be identified with the space $\mathcal{H}_{0}$ supporting the fundamental representation $\pi_{0}$ of the groupoid defined below.
Moreover, we check easily that $\pi_{\rho}(f)\Psi_{g}=\Psi_{f\star g}$ and the fundamental state $|0\rangle_{\rho}$ corresponds to the ``wave function'' $\Psi_{\mathbf{1}}=\sum_{x\in\Omega} p_{x}|x\rangle$, with $p_{x}=\phi_{\rho}(\mathbf{1}_{x})=\rho(\mathbf{1}_{x})$.

The fundamental representation $\pi_{0}$ is defined as follows: the supporting space is $\mathcal{H}_{0}=\sum_{x\in\Omega}\mathbb{C}|x\rangle$, and $\pi_{0}(f)|x\rangle=\sum_{\alpha\colon x\rightarrow y}f(\alpha)|y\rangle$, with $f=\sum_{\alpha\in \G}f(\alpha)\delta_{\alpha}$.
Then, clearly, we have
\be
\langle y|\pi_{0}(f)|x\rangle=\sum_{\alpha\colon x\rightarrow y} f(\alpha).
\ee
In particular, if we take our observable to be the characteristic function $\phi_{\rho}$ of the state $\rho$, then
\be
\langle y|\pi_{0}(\phi)|x\rangle=\sum_{\alpha\colon x\rightarrow y}\phi_{\rho}(\alpha)
\ee
and we can interpret this number as the probability amplitude of the outcome $y$ being returned after the outcome $x$. Using a more compact notation $\phi_{xy}$ for the amplitude $\langle y|\pi_{0}(\phi)|x\rangle$, if $\rho$ is a factorizable state, then we get easily the reproducibilty property for the amplitudes \cite{Ib18c} 
\be\label{eqn: reproducibility}
\phi_{zx}=\sum_{y\in\Omega}\phi_{zy}\,\phi_{yx}\,.
\ee

Finally, we will introduce a quantum measure associated with the state $\rho$ (see \cite{Ib18c}). 
First, we realize that the state $\rho$ on $C^{*}(\G)$ defines a decoherence functional $D$ on the $\sigma$-algebra $\Sigma$ of events of the groupoid $\G$ by means of 
\be
D(A,B):=\sum_{\alpha\in A,\beta\in B} \phi_{\rho}(\alpha^{-1}\circ \beta) \, , 
\ee
where $\alpha$ and $\beta$ must be composable, and, according to R. Sorkin's statistical interpretation of quantum systems (see \cite{So94,So95}), we define a quantum measure $\mu$ on $\Sigma$ as:
\be
\mu(A):=D(A,A),
\ee
where the grade 2-measure $\mu$ is an indicator of interference phenomena and measure zero events  $A\in\Sigma$, i.e.,  $\mu (A) = 0$, are physically precluded.
In particular, we can consider the event $A_{y,x}\subset \G$ defined as the set of transitions that provide the outcome $y$ after the outcome $x$, or, more simply stated, ``of obtaining $y$ after $x$''. Notice that using groupoids notations (see \cite{Ib18}), $A_{y,x} = G_+(x) \cap G_-(y) = G(y,x) = \{\alpha\in\G|\alpha:x\rightarrow y\}$, where $G_+(x) = \{ \alpha \colon x \to y\}$ and $G_-(y) = \{ \alpha \colon x \to y\}$. Then, we get:
\begin{eqnarray}
\label{eqn: statistical significance}
\mu(A_{y,x})&=D(A_{y,x},A_{y,x})=\sum_{\alpha,\beta\in\G(y,x)}\,\phi_{\rho}(\alpha^{-1}\circ\beta)=  \sum_{\alpha,\beta\in\G(y,x)}\,\overline{\phi_{\rho}(\alpha)}\,\phi_{\rho}(\beta)= \\
&=\overline{\sum_{\alpha\in\G(y,x)}\phi_{\rho}(\alpha)}\, \sum_{\beta\in\G(y,x)}\phi_{\rho}(\beta) = |\phi_{yx}|^{2} \, ,
\end{eqnarray}
which matches perfectly Feynman's prescription for the properties of amplitudes: reproducibility (Eq.\eqref{eqn: reproducibility}), and statistical significance, i.e., the module square of the amplitudes $\phi_{yx}$ define the statistical probability of the event (Eq.\eqref{eqn: statistical significance}).

\section{Evolution}

Evolution is described easily by the adaptation of Heisenberg's picture to the formalism we are describing here.
Thus, evolution will be described by a one-parameter group $\Phi_{t}$ of automorphisms\footnote{This condition can be weakened to consider positive or completely-positive maps.} of the $C^{*}$-algebra of the groupoid $\G$.
Given an observable $a$, its evolution will be given by the family $a_{t}=\Phi_{t}(a)$.
If the system is further determined by the state $\rho$, the expected values of the observable will change with time according to $\langle a_{t}\rangle_{\rho}=\rho(a_{t})=\rho(\Phi_{t}(a))$. 
Alternatively, we may think that the state $\rho$ changes with time as $\rho_{t}(a)=\rho(a_{t})$ instead, even if we will use the first interpretation.

Under appropriate regularity conditions, the one-parameter group $\Phi_{t}$ determines a derivation (in general just densely defined) of the algebra $C^{*}(\G)$ as:
\begin{equation}
D(a)=\left.\frac{\mathrm{d}}{\mathrm{d}t}\Phi_{t}(a)\right|_{t=0},
\end{equation}
and, provided that the algebra $C^{*}(\G)$ is semisimple (which is the case for finite groupoids \cite{Ib18d}), then there will exist an element $h\in C^{*}(\G)$ such that 
\begin{equation}
D(a)=\imath[a,h],
\end{equation}
and that will be called, accordingly, the Hamiltonian of the system.
Then, we obtain easily that
\begin{equation}
\Phi_{t}(a)=u_{t}^{\dagger}\,a\,u_{t},
\end{equation}
with $u_{t}=\mathrm{exp}(\imath t h)$ a one-parameter subgroup of unitary elements in $C^{*}(\G)$.
We may write the evolution equation in Heisenberg's form as:
\begin{equation}\label{eqn: Heisenberg's equation}
\frac{\mathrm{d}}{\mathrm{d}t}a=\imath[a,h].
\end{equation}
The state $\rho$ allows for a natural representation of the evolution equation \eqref{eqn: Heisenberg's equation} in the GNS Hilbert space $\mathcal{H}_{\rho}$.
Indeed, since  $\rho(a)=\langle 0|\pi_{\rho}(a)|0\rangle$, we get that 
\begin{equation}
\langle a_{t}\rangle_{\rho}=\rho(a_{t})=\langle 0|\pi_{\rho}(u_{t})^{\dagger}\,\pi_{\rho}(a)\,\pi_{\rho}(u_{t})|0\rangle= \langle \psi_{t}|\pi_{\rho}(a)|\psi_{t}\rangle
\end{equation}
with $|\psi_{t} \rangle =\pi_{\rho}(u_{t})|0\rangle=\mathrm{U}_{t}|0\rangle$,  with $\mathrm{U}_{t}$ the one-parameter group of unitary operators on $\mathcal{H}_{\rho}$ associated with the unitary elements $u_{t}\in C^{*}(\G)$ by the GNS representation.
Note that 
\begin{equation}
\pi_{\rho}(h)|\psi_{t}\rangle = - \imath \frac{\mathrm{d}}{\mathrm{d}t} \pi_\rho (\exp (\imath t h)) |0 \rangle = - \imath  \frac{\mathrm{d}}{\mathrm{d}t}\mathrm{U}_{t}|0\rangle=  H |\psi_{t}\rangle \,,
\end{equation} 
then, $H=\pi_{\rho}(h)$ is a Hamiltonian operator on $\mathcal{H}_{\rho}$, and Schr\"{o}dinger's equation:
\begin{equation}
\imath\,\frac{\mathrm{d}}{\mathrm{d}t}|\psi\rangle= H|\psi\rangle \, ,
\end{equation}

can be considered as the GNS representation of the evolution equation \eqref{eqn: Heisenberg's equation}.

In the particular instance of $\rho$ being a factorizable state, we may use $| \Psi_{t}\rangle=\mathrm{U}_{t}|0\rangle$ as wave function, or, even better, we may use Feynman's wave function
\begin{equation}
|\psi_{F}\rangle=\pi_{\rho}(\mathrm{I})|0\rangle,
\end{equation}
where $\mathbb{I}$ is the ``incidence matrix'' element in the $C^*$-algebra of the groupoid $\G$:
\begin{equation}
\mathbb{I}=\sum_{\alpha\in \G}\,\alpha\,.
\end{equation}
Therefore, we obtain
\begin{equation}
\psi_{F}(x)=\sum_{\alpha\in\G_{+}(x)}\,\mathrm{e}^{\imath s(\alpha)}\,.
\end{equation}

\section{The quantum ratchet}

Let us discuss a single example that could help to illustrate the abstract background provided by the groupoid interpretation of Schwinger's algebra.
The system that we will be considering will just have two outputs, say $+$ and $-$.
However, contrary to what happens with the standard q-bit (see \cite{Ib18b}), there are more than one transition between the outcomes $+$ and $-$.

Imagine that the system has an inner register which is not available to our scrutiny, and such that each time a transition happens, the register changes in a well-defined way.
For instance, the register could consist in a finite list of symbols, and each time there happens a transition, these symbols are permuted in a fixed way.
A simple instance we may consider is when the inner register has 3 symbols, and the permutation is a cycle $\sigma$  (for instance the cycle $\sigma=(123)$).
Thus, our system will be a ``ratchet'' such that each time a transition $+\rightarrow -$ is detected the inner register is permuted (always with the same permutation $\sigma$).
In consequence, there will be two elementary transitions $\alpha_{1}$ (from $-$ to $+$ and $\sigma$ in the register), and $\beta_{1}$ (from $+$ to $-$ and $\sigma$ in the inner register).

The groupoid generated by this quiver $Q=\{\alpha_{1},\beta_{1}\}\rightrightarrows\{+,-\}$ will have $12$ elements which are listed below.
We will use the notation $(y, \pi,x)$ to indicate the transtion that takes us from $x$ to $y$ ($x,y=\pm$) and that acts in the inner register with the permutation $\pi=e,\sigma,\sigma^{2}$:
\begin{equation}
\begin{array}{ccccccc}
\alpha_{1}=(+,\sigma,-)& &\beta_{1}=(-,\sigma,+) & &\mathbf{1}_{+}=(+,e,+) & &\mathbf{1}_{-}=(-,e,-) \\
\alpha_{2}=(+,\sigma^{2},-)& &\beta_{2}=(-,\sigma^{2},+) & &\sigma_{+}=(+,\sigma,+) & & \sigma_{-}=(-,\sigma,-) \\ 
\alpha_{3}=(+,e,-)& &\beta_{3}=(-,e,+) & &\sigma_{+}^{2}=(+,\sigma^{2},+) & &\sigma_{-}^{2}=(-,\sigma^{2},-) 
\end{array}
\end{equation}

This groupoid, whose multiplication table is displayed below (see Table \ref{mul_table}), is the natural groupoid extension of the notion of cyclic groups and could be properly called a cyclic groupoid, in this case the cyclic groupoid $C_{2,3}$ (see \cite{Ib19}).
\begin{table}[h!]
\centering

\begin{tabular}{|c|c c c|c c c|c c c|c c c|}
\hline
$(C_{2,3}, \circ)$ & $\1_+$ & $\sigma_+$ & $\sigma^2_+$ & $\alpha_1$ & $\alpha_2$ & $\alpha_3$ & $\1_-$ & $\sigma_-$ & $\sigma_-^2$ & $\beta_1$ & $\beta_2$ & $\beta_3$ \\
\hline
$\1_+$ & $\1_+$ & $\sigma_+$ & $\sigma^2_+$ & $\alpha_1$ & $\alpha_2$ & $\alpha_3$ & * & * & * & * & * & * \\
$\sigma_+$ & $\sigma_+$ & $\sigma^2_+$ & $\1_+$ & $\alpha_2$ & $\alpha_3$ & $\alpha_1$ & * &* & * &* &* & * \\
$\sigma^2_+$ & $\sigma^2_+$ & $\1_+$ & $\sigma_+$ & $\alpha_3$ & $\alpha_1$ & $\alpha_2$ & * & * & * & * & * & * \\
\hline
$\beta_1$ & $\beta_1$ & $\beta_2$ & $\beta_3$ & $\sigma_-^2$ & $\1_-$ & $\sigma_-$ & * & * & * & * & * & * \\
$\beta_2$ & $\beta_2$ & $\beta_3$ & $\beta_1$ & $\1_-$ & $\sigma_-$ & $\sigma_-^2$ &  * & * & * & * & * & * \\
$\beta_3$ & $\beta_3$ & $\beta_1$ & $\beta_2$ & $\sigma_-$ & $\sigma_-^2$ & $\1_-$ & * & * & * & * & * & * \\
\hline
$\1_-$ & * & * & * & * & * & * & $\1_-$ & $\sigma_-$ & $\sigma_-^2$ & $\beta_1$ & $\beta_2$ & $\beta_3$ \\
$\sigma_-$ & * & * & * & * & * & * & $\sigma_-$ & $\sigma_-^2$ & $\1_-$ & $\beta_2$ & $\beta_3$ & $\beta_1$ \\
$\sigma_-^2$ & * & * & * & * & * & * & $\sigma_-^2$ & $\1_-$ & $\sigma_-$ & $\beta_3$ & $\beta_1$ & $\beta_2$ \\
\hline
$\alpha_1$ & * & * & * & * & * & *&  $\alpha_1$ & $\alpha_2$ & $\alpha_3$ & $\sigma^2_+$ & $\1_+$ & $\sigma_+$ \\
$\alpha_2$ & * & * & * & * & * & *& $\alpha_2$ & $\alpha_3$ & $\alpha_1$ & $\1_+$ & $\sigma_+$ & $\sigma^2_+$ \\
$\alpha_3$ & * & * & * & * & * & *&   $\alpha_3$ & $\alpha_1$ & $\alpha_2$ & $\sigma_+$ & $\sigma^2_+$ & $\1_+$ \\ 
\hline
\end{tabular}
\caption{Multiplication table of the cyclic groupoid $C_{2,3}$.}\label{mul_table}
\end{table}
We may specify a factorizable state $\rho$ on $C_{2,3}$ by stating the values of its associated function $\varphi$ on the elementary transitions $\alpha_{1},\beta_{1}$.
Thus, we may define:
\begin{eqnarray*}
\varphi(\alpha_{1})& = &\mathrm{e}^{\imath s} \\
\varphi(\beta_{1})& = & \mathrm{e}^{\imath(\delta - s)}=\mathrm{e}^{\imath\delta}\,\overline{\varphi(\alpha_{1})},\;\;\delta\in\mathbb{R}\,.
\end{eqnarray*}
This implies that the ratchet produces a phase shift $\delta$ when ``moving back''.
Then, using the multiplication table we get easily:
\begin{equation}
\begin{array}{ccccccc}
\varphi(\alpha_{1})=\mathrm{e}^{\imath s} & \;\;&\varphi(\beta_{1})=\mathrm{e}^{\imath(\delta -s)} & \;\;&\varphi(\mathbf{1}_{+})=\mathrm{e}^{3\imath \delta} & \;\;&\varphi(\mathbf{1}_{-})= \mathrm{e}^{-3\imath \delta} \\
\varphi(\alpha_{2})=\mathrm{e}^{\imath s}\mathrm{e}^{2\imath\delta} &\;\;& \varphi(\beta_{2})=\mathrm{e}^{-\imath s} & \;\;&\varphi(\mathbf{\sigma}_{+})=\mathrm{e}^{2\imath \delta} & \;\;& \varphi(\mathbf{\sigma}_{-})= \mathrm{e}^{-2\imath \delta} \\
\varphi(\alpha_{3})=\mathrm{e}^{\imath s}\mathrm{e}^{3\imath \delta} &\;\;& \varphi(\beta_{3})=\mathrm{e}^{-\imath s}\mathrm{e}^{-\imath \delta} & \;\;& \varphi(\mathbf{\sigma}_{+}^{2})=\mathrm{e}^{\imath \delta} & \;\;& \varphi(\mathbf{\sigma}_{-}^{2})= \mathrm{e}^{-\imath \delta} .
\end{array}
\end{equation}
Recalling that $\varphi$ is a function generating a factorizable state satisfying the unitarity property (see the end of section \ref{sec: observable and states}), we have the constraint $3\delta=2\pi$.
The GNS representation of the state $\rho$ will have support $\mathcal{H}_{\rho}\cong\mathbb{C}^{2}$, with basis $|+\rangle,|-\rangle$.

Let us now consider the hamiltonian function $h \, : \, C_{2,3}\, \rightarrow \, \mathbb{C}$ written as
\begin{equation}
h = \delta_{\alpha_1} + \delta_{\beta_2} + \delta_{\alpha_2} + \delta_{\beta_1} +\delta_{\alpha_3} + \delta_{\beta_3}\,.
\end{equation}
The associated one parameter group of unitary transformations $u_t$ belonging to the groupoid-algebra $C^*(C_{2,3})$ is expressed as follows:
\begin{equation}
u_t = \frac{1}{3}(\cos(3t)-1) \left(\delta_{\1_+} + \delta_{\sigma_+} + \delta_{\sigma_+^2} + \delta_{\1_-} + \delta_{\sigma_-} + \delta_{\sigma_-^2}  \right) + \frac{\imath}{3} \sin(3t) h + \delta_{\1_+} + \delta_{\1_-}\,. 
\end{equation} 

The following computations provide  some insight into the physical interpretation of this ``quantum ratchet''. 
Indeed, we can compute the probability amplitude $\left\langle \psi_{\1_+} | \psi_{\1_+}(t) \right\rangle $ to get:
\begin{eqnarray*}
&\left\langle \psi_{\1_+} | \psi_{\1_+}(t) \right\rangle  =  \rho(\1_+ u_t \1_+) = \rho\left( \frac{1}{3}(\cos(3t)-1)(\delta_{\1_+} +\delta_{\sigma_+} + \delta_{\sigma_+^2}) + \delta_{\1_+} \right) = \\
& = \frac{1}{2} \left( 1+\frac{1}{3} (\cos(3t)-1)(1+2\cos(\delta)) \right) = \frac{1}{2}\,,
\end{eqnarray*}
and a similar computation shows that
\be
\left\langle \psi_{\1_+} | \psi_{\1_+}(t) \right\rangle = \left\langle \psi_{\1_-} | \psi_{\1_-}(t) \right\rangle .
\ee
Analogously, we have
\begin{eqnarray*}
& \left\langle \psi_{\1_-} | \psi_{\1_+}(t) \right\rangle =  \rho (\1_- u_t \1_+) = \rho \left( \frac{\imath}{3} \sin(3t) (\delta_{\beta_1} + \delta_{\beta_2} + \delta_{\beta_3}) \right) = \\
& =  \frac{\imath}{6} \sin(3t)\left( \mathrm{e}^{-\imath S }(1+\cos(2\delta)) \right)=0,
\end{eqnarray*}
and
\begin{equation}
\left\langle \psi_{\1_-} | \psi_{\1_+}(t) \right\rangle=\left\langle \psi_{\1_+} | \psi_{\1_{-}}(t) \right\rangle .
\end{equation}
These values  result from the interference of the transitions involved. In particular, the transitions due to the internal degrees of freedom produce a destructive interference according to which the previous probability amplitudes are constant in time, a result which is very different from the ``simple'' qubit case. 

Nevertheless, the qubit results are recovered when one considers the Hamiltonian function $\tilde{h} = \frac{1}{2}(\delta_{\alpha_2}+\delta_{\beta_1})$ which generates the following one parameter group of unitary transformations
\begin{equation}
\tilde{u}_t = \cos\left( \frac{t}{2} \right) (\delta_{\1_+} + \delta_{\1_-}) + \imath \sin\left( \frac{t}{2} \right) \left( \delta_{\alpha_2} + \delta_{\beta_1} \right)\,.
\end{equation} 

According to this evolution, the previous probability amplitudes are different since we get
\begin{eqnarray*}
&\left\langle \psi_{\1_+}| \tilde{u}_t | \psi_{\1_+} \right\rangle = \left\langle \psi_{\1_-}| \tilde{u}_t | \psi_{\1_-} \right\rangle = \frac{1}{2}\cos\left( \frac{t}{2} \right)\\
&\left\langle \psi_{\1_-}| \tilde{u}_t | \psi_{\1_+} \right\rangle = -\overline{\left\langle \psi_{\1_+}| \tilde{u}_t | \psi_{\1_-} \right\rangle} = \frac{\imath}{2}\sin\left( \frac{t}{2} \right)\mathrm{e}^{\imath (S-\delta)}\,,
\end{eqnarray*}
which coincides with the probability amplitudes of a qubit, except for the $\delta$ factor in the exponential.


\section*{Acknowledgments}

F.D.C. and A.I. would like to thank partial support provided by the MINECO research project MTM2017-84098-P and QUITEMAD++, S2018/TCS-A4342. A.I. and G.M. acknowledge financial support from the Spanish Ministry of Economy and Competitiveness, through the Severo Ochoa Programme for Centres of Excellence in RD(SEV-2015/0554). G.M. would like to thank the support provided by the Santander/UC3M Excellence Chair Programme 2019/2020, and he is also a member of the Gruppo Nazionale di Fisica Matematica (INDAM), Italy.



\end{document}